\documentclass{sigchi}

\doi{}
\isbn{}
\usepackage{balance}       
\usepackage{graphics}      
\usepackage[T1]{fontenc}   
\usepackage{txfonts}
\usepackage{mathptmx}
\usepackage[pdflang={en-US},pdftex]{hyperref}
\usepackage{color}
\usepackage{booktabs}
\usepackage{textcomp}
\usepackage{microtype}        
\usepackage{ccicons}          
\usepackage{todonotes}
\usepackage{xspace}

\newcommand{\meansd}[2]{{$(M=#1$, $SD=#2)$}} \newcommand{\wilcoxon}[2]{($Z=#1$, $p < #2$)}

\def\approachname{our approach\xspace}
\def\plaintitle{DialPlate: Enhancing the Detection of Smooth Pursuits Eye Movements Using Linear Regression}

\def\emptyauthor{}
\def\plainkeywords{Eye tracking; Smooth pursuits; Distinguishable targets}

\makeatletter
\def\url@leostyle{%
  \@ifundefined{selectfont}{
    \def\UrlFont{\sf}
  }{
    \def\UrlFont{\small\bf\ttfamily}
  }}
\makeatother
\def\pprw{8.5in}
\def\pprh{11in}

\definecolor{linkColor}{RGB}{6,125,233}
\hypersetup{
  pdftitle={\plaintitle},
  pdfauthor={\emptyauthor},
  pdfkeywords={\plainkeywords},
  pdfdisplaydoctitle=true, 
  bookmarksnumbered,
  pdfstartview={FitH},
  colorlinks,
  citecolor=black,
  filecolor=black,
  linkcolor=black,
  urlcolor=linkColor,
  breaklinks=true,
  hypertexnames=false
}

\begin{document}

\title{\plaintitle}

\numberofauthors{3}
\author{
  \alignauthor{Heiko Drewes\\
    \affaddr{University of Munich LMU}\\
    \affaddr{Munich, Germany}\\
    \email{heiko.drewes@ifi.lmu.de}}\\
  \alignauthor{Mohamed Khamis\\
    \affaddr{University of Munich LMU}\\
    \affaddr{Munich, Germany}\\
    \email{mohamed.khamis@ifi.lmu.de}}\\
  \alignauthor{Florian Alt\\
    \affaddr{University of Munich LMU}\\
    \affaddr{Munich, Germany}\\
    \email{florian.alt@ifi.lmu.de}}\\
}

\maketitle

\begin{abstract}
We introduce and evaluate a novel approach for detecting smooth pursuit eye movements that increases the number of distinguishable targets and is more robust against false positives. 
Being natural and calibration-free, Pursuits has been gaining popularity in the past years. 
At the same time, current implementations show poor performance when more than eight on-screen targets are being used, thus limiting its applicability. 
Our approach (1) leverages the slope of a regression line, and (2) introduces a minimum signal duration that improves both the new and the traditional detection method. 
After introducing the approach as well as the implementation, we compare it to the traditional correlation-based Pursuits detection method. 
We tested the approach up to 24 targets and show that, if accepting a similar error rate, nearly twice as many targets can be distinguished compared to state of the art. For fewer targets, accuracy increases significantly. 
We believe our approach will enable more robust pursuit-based user interfaces, thus making it valuable for both researchers and practitioners.
\end{abstract}

\vspace{-1mm}
\category{H.5.2.}{Information Interfaces \& Presentation}{User Interfaces -- Input Devices and Strategies}
\vspace{-1mm}

\section{Introduction}

Since the early 80s there is a vision that gaze-based interfaces could make our interaction with computer easier and more efficient  \cite{Bolt:1981}. 
Gaze-based interfaces have many promises: they work over distances, they are hygienic as there is nothing to touch, they keep the hands free for other tasks, they are silent, and they are maintenance-free as eye trackers have no moving parts. 
At the same time, gaze-based interfaces usually need a time consuming calibration, they lack high accuracy, and they are prone to the so-called Midas touch problem \cite{Jacob:1990}.

In 2013, Vidal et al.~introduced a novel concept for gaze interaction based on smooth pursuit eye movements \cite{Vidal:2013}. 
In interfaces with moving targets, they compare the user's gaze and the movement of the target, hence allowing a matching pursuit movement to be detected by calculating Pearson's correlation coefficient. 
The strength of this approach is its independence from offset and scaling and, therefore, the eye tracker does not need to be calibrated but can be instantly used. 
Another advantage is that due to being independent from scale, interfaces on small areas, such as a smartwatch display \cite{Esteves:2015}, can be built.
A typical interface based on smooth pursuits offers several targets to give the user a choice. 
Esteves et al.~\cite{Esteves:2015} showed that it is possible to distinguish eight targets moving on a circle. 
However, they reported false positive rates of 12\% for pursuits-based interaction. 
We argue that to make gaze-based interaction usable in everyday life, this rate needs to be significantly reduced. 
Similarly, Vidal et al.~showed that detection accuracy drops significantly when showing more than 8 targets moving at the same speed and trajectory \cite{Esteves:2015}. 

\begin{figure} [t]
\centering
  \includegraphics[width=0.45\textwidth]{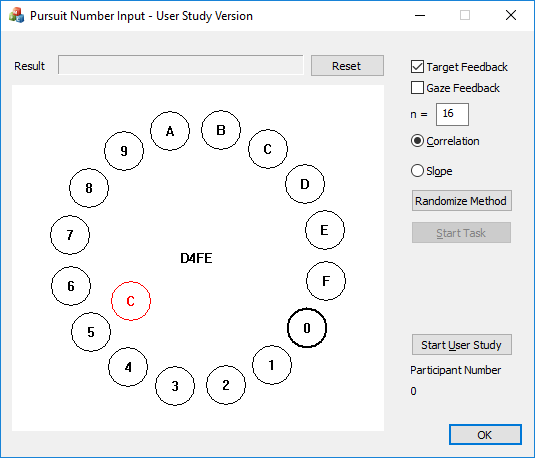}
  \caption{We present an approach to enhance the detection of smooth pursuits eye movement. In particular, by using the slope of a regression line, our approach allows for (a) increasing the number of distinguishable targets and (b) decreasing the number of false positives. }
 \label{figure:screenshotPNI}
 \vspace{-3mm}
\end{figure}

This underpins an inherent challenge in Pursuits-based interfaces -- the number of targets and reliability present a trade-off: reducing the number of targets increases the detection reliability and vice versa. 
At the same time, today's interfaces provide many different elements, such as the number of application icons on a smartphone or the keys on a soft keyboard. 

To address this, we introduce a new Pursuits detection method that increases the accuracy of selections even with high numbers of on-screen targets. 
Rather than the widely used Pearson correlation \cite{Esteves:2015,Esteves:2017:SSP:3126594.3126616,Khamis:2015:FSS:2800835.2804335,Khamis:2017:EAE:3126594.3126630,khamis2018avi,Khamis:2016:TUT:2971648.2971679,Khamis:2016:EWU:3012709.3012743,Velloso:2017:MCS:3086563.3064937,Velloso:2016:ADC:2901790.2901867}, our novel method uses the slope of a regression line. 

In a study (N=16), we compared the performance of \approachname to the state-of-the-art for pursuits-based interfaces. 
In particular, we compared the influence of the number of targets on input speed and error rate. 
Results show that \approachname allows up to 24 targets to be distinguished. 
For eight or more targets it reduces the error rate by factor 5 to 10 compared to the state-of-the-art approach. 
We built a sample application and discuss how \approachname supports designers in building highly reliable calibration-free gaze-based interfaces. 

The contribution of this work is twofold: First, we describe a novel detection method for smooth pursuits eye movements.  Second, we report on a comparison of the approach with state-of-the-art, revealing a significant increase in number of detectable targets as well as in accuracy.

\vspace{-1mm}
\section{Background and Related Work}

While early works in gaze-based interaction relied mostly on fixations, the research community started to move towards detecting gaze behavior, such as gaze gestures \cite{Drewes2007}, and more recently smooth pursuit \cite{Vidal:2013}. 
Smooth pursuit eye movements are naturally performed when gazing at a moving target. 
Interaction using smooth pursuit (aka Pursuits) is promising since it does not require calibration because it relies on relative eye movements rather than precise fixation points. 

\vspace{-1mm}
\subsection{Applications of Pursuits}

Pursuits has been utilized in several applications and domains.
Being a calibration-free and contactless gaze-only modality, a large body of work investigated its use on public displays, where immediate usability is essential. 
For example, Vidal et al.~used Pursuits on public displays for gaming and entertainment applications \cite{Vidal:2013}. 
In EyeVote, Pursuits was used for voting on public displays~\cite{Khamis:2016:EWU:3012709.3012743}. 
Pursuits was also successfully deployed in active eye tracking settings, where the tracker moved on a rail system to follow users as they pass by large public displays \cite{Khamis:2017:EAE:3126594.3126630}. 
Lutz et al.~used Pursuits for entering text on public displays \cite{JEMR2394}. 
They worked around Pursuits' limitations by performing each letter's selection on two stages: the user first selects one of 5 groups of letters, the group then expands to allow the user to finally select the desired letter. 

Other ubiquitous technologies leveraged Pursuits as well. 
Esteves et al.~\cite{Esteves:2015} used Pursuits for gaze interaction with smart watches. 
Velloso et al.~\cite{Velloso:2016} utilized Pursuits in smart homes.

Pursuits were also used in mixed reality. VR benefits from using Pursuits during interaction, especially when moving in VR \cite{khamis2018avi}, and when interacting with occluded targets \cite{7893315}. Pursuits was also employed in augmented reality glasses \cite{Esteves:2017:SSP:3126594.3126616}. 

Eye movements are subtle and hard to observe. Hence Pursuits was used for authentication~\cite{Cymek:2014,Rajanna:2018,Rajanna:2017:GGU:3027063.3053070}. 

As for desktop settings, Kangas et al.~\cite{Kangas:2016} and \v{S}pakov et al.~\cite{Spakov:2016} employed Pursuits in the form of a  continuous signal to control on-screen widget to, for example, adjust volume. 

In addition to using it as a calibration-free gaze interaction technique, Pursuits can also be used for calibration. Pfeuffer et al.~\cite{Pfeuffer:2013:PCM:2501988.2501998} introduced a method to calibrate the eye tracker as users follow on-screen moving targets. Similarly, Celebi et al.~\cite{Celebi:2014:SPC:2578153.2583042} used Pursuits for eye tracker calibration. 
Khamis et al.~\cite{Khamis:2016:TUT:2971648.2971679} used gradually revealing text to calibrate the eye tracker while users read-and-pursue. A major drawback of previous works is that the interfaces often has a limited number of targets shown at once. 
Previous implementations could distinguish up to 8 targets reliably \cite{Esteves:2015,Velloso:2017:MCS:3086563.3064937,Vidal:2013}. We show that it is possible to distinguish 24 targets with significantly higher accuracy compared to state of the art. 

\subsection{Implementations of Pursuits}

There are two predominant implementations of Pursuits detection for interaction, one of which uses the Euclidean distance between the gaze estimates and target positions \cite{Kangas:2016,Rajanna:2018,Rajanna:2017:GGU:3027063.3053070,Spakov:2016}, while the other one employ Pearson's product moment correlation \cite{Esteves:2015,Esteves:2017:SSP:3126594.3126616,Khamis:2015:FSS:2800835.2804335,Khamis:2017:EAE:3126594.3126630,khamis2018avi,Khamis:2016:TUT:2971648.2971679,Khamis:2016:EWU:3012709.3012743,Velloso:2017:MCS:3086563.3064937,Velloso:2016:ADC:2901790.2901867}. 

The Euclidean distance method is susceptible to inaccurate detection in the presence of an offset between the real gaze point and the estimated one. This means that it is not reliable when the eye tracker is not calibrated, or when the gaze estimation is not accurate. In contrast, the correlation method is independent of offsets and scaling. For this reason, it works reliably without calibration \cite{Khamis:2016:EWU:3012709.3012743,Velloso:2017:MCS:3086563.3064937,Vidal:2013}, and even on small interfaces such that of smart watches \cite{Esteves:2015}. On the downside, the accuracy of the correlation-based detection drops significantly in the presence of more than 8 targets \cite{Esteves:2015,Velloso:2017:MCS:3086563.3064937,Vidal:2013}.

\section{Regression Slope-based Detection of Pursuits}

We introduce a novel approach of detecting Pursuits and start with theoretical foundations before describing our enhancements and implementation.

\subsection{Theoretical Background} \label{Theoretical Considerations}
A smooth pursuit detection algorithm receives the gaze coordinates and the coordinates of the on-screen targets as input. 
It collects a certain number of data samples, calculates a metric function for each target, and then compares the metric values of each target with a threshold or threshold interval. Targets whose metric values match the threshold condition are reported as detected. Typical metric functions for pursuit detection are Euclidean distance \cite{Kangas:2016,Rajanna:2018,Rajanna:2017:GGU:3027063.3053070,Spakov:2016} or correlation \cite{Esteves:2015,Esteves:2017:SSP:3126594.3126616,Khamis:2015:FSS:2800835.2804335,Khamis:2017:EAE:3126594.3126630,khamis2018avi,Khamis:2016:TUT:2971648.2971679,Khamis:2016:EWU:3012709.3012743,Velloso:2017:MCS:3086563.3064937,Velloso:2016:ADC:2901790.2901867}. Detection algorithms using Euclidean distance need a calibrated eye tracker \cite{Rajanna:2018,Rajanna:2017:GGU:3027063.3053070,Spakov:2016}, while detection methods using correlation are independent from offset and scaling \cite{Velloso:2017:MCS:3086563.3064937,Vidal:2013}. The implicit assumption behind this statement is that the calibration error can be described by an affine transformation. 

\subsection{From Correlation-based to Slope-based detection}

The algorithm described here works with linear regression which is, in terms of mathematics, closely related to correlation. 
Linear regression and correlation need a list of value pairs which in our case are the x-coordinates of the gaze g\textsubscript{x} and the target position t\textsubscript{x} or the y-coordinates, respectively. Every value pair can be plotted in a plane. The linear regression analysis finds the straight line that best fits the plotted data. 

The regression coefficient is the \emph{slope} of the line, the \emph{intercept} is the value where the line crosses the abscissa and the \emph{correlation} is a measure for the quality of the fit. If the gaze follows the target perfectly and the eye tracker provides accurate positions, then g\textsubscript{x} = t\textsubscript{x} and the plot is a bisecting line of ordinate and abscissa with intercept=0.0, slope=1.0, and correlation=1.0. If the gaze \emph{does not follow the target}, the values for intercept, slope and correlation are very different from the perfect values.  

The correlation detection method typically requires a correlation value higher than 0.8 \cite{Velloso:2017:MCS:3086563.3064937}. A \emph{calibration error} results in an intercept (=offset) different from zero and an only slightly changed value for the slope (=scaling factor) while the correlation does not change. Our pilot studies showed that calibration errors for the scaling factor are in a range from 0.9 to 1.1. 

\vspace{-2mm}
\subsection{Advantages of Slope-based Pursuits detection}

Our method presented here requires the slope to be close to 1.0 -- hence, we refer to this method as slope detection. 
For the study, we used a threshold interval from 0.77 to 1.3. 

Similar to the correlation method, the slope is independent from offsets. Consequently, the slope method  detects Pursuits without calibration.

The slope detection has a further advantage: It distinguishes between synchronously moving targets of different trajectory sizes, while the correlation method does not. The reason is that the correlation is insensitive to offsets and scaling, while the regression line's slope is only insensitive to offsets. 

\vspace{-2mm}
\subsection{Implementation}
We implemented both detection methods, the correlation and the slope method. We used the following formulas:

Regression analysis:
\vspace{-3mm}
\[ s = \frac{n \displaystyle\sum_{i=1}^n x y - \displaystyle\sum_{i=1}^n x \displaystyle\sum_{i=1}^n y} { n \displaystyle\sum_{i=1}^n x^2 - (\displaystyle\sum_{i=1}^n x)^2} \]

Correlation:
\vspace{-3mm}
\[ r = \frac{n \displaystyle\sum_{i=1}^n x y - \displaystyle\sum_{i=1}^n x \displaystyle\sum_{i=1}^n y} { \sqrt{n \displaystyle\sum_{i=1}^n x^2   - (\displaystyle\sum_{i=1}^n x)^2 } \sqrt{n \displaystyle\sum_{i=1}^n y^2   - (\displaystyle\sum_{i=1}^n y)^2 } } \]
\\
where x is a gaze coordinate, y the corresponding target coordinate, and n the size of the data window.

In contrast to formulas which require mean values and consequently need to sum up values over all data in the data window, these formulas allow a sliding window by only subtracting an old value and adding a new value. 
As a result, the algorithm's run time is independent from the data window size. 

We further enhanced the algorithm. For a positive detection, rather than relying on a single sample as in previous work \cite{Velloso:2017:MCS:3086563.3064937,Vidal:2013}, our threshold condition needs to be met for a number of consecutive samples, hence introducing a \emph{minimum signal duration}. The minimum signal duration reduces false positives. 
Reducing false positives is also possible by increasing the data window size. However, pilot studies showed that a small data window and a minimum signal duration excludes more false positives compared to a larger data window. 

As a further enhancement, we added some \emph{smoothing} to the gaze signal by calculating the average over the last k samples. Smoothing the gaze signal improves the detection with the slope method but increases the false positive rate for the correlation detection. For a fair comparison in the user study, we  used the smoothed signal only for the slope method. We also adjusted the minimum signal duration for best results.

While pilot testing, we observed that a false positive detection of the same target often followed successful detections (despite clearing all buffers after a successful detection). We found the reason to be the reaction time of the user who usually continues gazing at the target after successful detection. To address this, we dropped some samples after a positive detection.

Table \ref{table:Parameters} shows the parameters used for both detection methods. We used an eye tracker which delivers 60 samples per second. 

\begin{table} [t]
\centering
\begin{tabular}{  l  r  r } 
 \toprule
 Parameter & Correlation Method & Slope Method\\ [0.5ex] 
 \midrule
 Window size & 30 samples & 30 samples\\ 
 Smoothing & 0 samples & 20 samples \\ 
 Minimum duration & 20 samples & 15 samples \\
 Threshold & 0.8 & 0.77 -- 1.3 \\
 Skipped samples & 30 samples & 30 samples \\
  \bottomrule
\end{tabular}
\caption{Parameters used for correlation and slope detection methods}
\vspace{-3mm}
\label{table:Parameters}
\end{table}

\section{Evaluation}
We conducted a user study to compare our approach to the state of the art method for detecting Pursuits.

\subsection{Apparatus}
To evaluate our Pursuits detection approach, we developed a sample application (see Figure \ref{figure:screenshotPNI}) in which users can enter digits (0 to 9) and letters (A to N) via Pursuits. 

The application runs on an Acer Aspire V17 Nitro laptop with integrated Tobii IS4 Base AC eye tracker (60 Hz). 
The display has a resolution of 1920 times 1080 pixels on 38.4\,cm times 21.7\,cm, which results in 0.2\,mm for one pixel or 50\,px per centimeter. The average distance between the participants' eyes to the display is around 50\,cm +/- 5\,cm, which corresponds to 0.02$^{\circ}$ per pixel or around 50\,px per degree. The targets move clockwise on a circle with a radius of 130\,px (2.6$^{\circ}$), except for the `cancel'-target which moves counter-clockwise on a circle with a radius of 80\,px (1.6$^{\circ}$). The radius of each target is 20\,px (0.4$^{\circ}$) and they move at 6.5$^{\circ}$/s (2.5 seconds per rotation). 

The interface provides visual and acoustic feedback for the detection. Every target that matches the threshold condition is filled with color, whose intensity increases the longer the threshold condition stays true, and reaches its maximum once the minimum signal duration is reached. Different beeps are used for correct and wrong entries.

\subsection{Study Design}
The study was designed as a repeated-measures experiment with two independent variables. The first was the Pursuits detection method, with two conditions: correlation-based detection (baseline) and slope-based detection (\approachname). The second was the number of targets; participants went through 10 blocks, with the first block showing 6 targets, and gradually incremented the number of targets by 2 up to 24 simultaneously moving targets. The order of methods was randomized. The task of the user was to enter 4 symbols in each block.

\vspace{-1mm}
\subsection{Procedure}
We invited 16 participants (3 females) with normal or corrected to normal vision aged between 24 and 58. After arriving at the lab, participants filled out a form with the demographic data and received a short introduction to the system (Figure \ref{figure:screenshotPNI}). To test how well the methods work for spontaneous gaze interaction, we did not calibrate the eye tracker for each participant. Instead, it was calibrated only once by one of the authors. The participants' task was to enter a four-digit number by following the clockwise rotating number targets using gaze. In case of entering a wrong digit, the participants had to delete it by selecting the the counter-clockwise rotating `cancel'-target. 

Participants first completed a training task with six targets in which they entered four symbols (digits and letters), and tried to cancel an entry. These entries were excluded from the analysis. Participants then went through the 10 blocks, each covering a number of targets (6,8,10,12,14,16,18,20,22,24) and consisting of two selection tasks (one per detection method). 

Every selection task had a timeout of 90 seconds. If a participant was not able to fulfill the task in time or wished to abort, the study continued with the other method until the participant failed. We concluded with a semi-structured interview. 

\vspace{-1mm}
\subsection{Results}
Apart from the qualitative feedback and observations, we logged the \textit{maximum number of targets} shown simultaneously from which participants could still perform successful selections. We further logged the \textit{errors}, which correspond to the number of times users canceled their input. We also logged the average \textit{task completion time}, which denotes the time taken to enter all 4 symbols correctly. Finally we logged the average \textit{entry time} for entering each symbol. 
 
 \vspace{-1mm}
\subsubsection{Interviews and Observations}

All participants understood immediately how to operate the system and how to enter the digits, but it seemed that they were at the beginning of a steep learning curve. Many saw the user study like a computer game and were highly ambitious to reach a high score. All participants reported that the task required a lot of focussing. All participants reported that they found the slope-based method more accurate and easier, some of them even mentioned their preference before being asked. 

\subsubsection{Maximum Selectable Targets}

We counted the maximum number of displayed targets from which participants were able to enter the four demanded symbols (see the bars in Figure \ref{figure:graph_times}). The slope detection approach outperformed the correlation detection method. 
Only one participant was able to select more targets with the latter.

A Wilcoxcon signed ranked test revealed that the slope detection method results in a significant increase of the number of displayed targets from which participants successfully made selections \wilcoxon{3.168}{0.01}. Using the correlation method, the maximum target number with which the participants were able to accomplish the task was between 10 and 24 \meansd{15.0}{3.7}. Using the slope method, the maximum target number was between 8 and 24 \meansd{21.6}{4.6}. 

\subsubsection{Errors}

\begin{figure} [t]
\centering
\includegraphics[trim={2cm 16cm 3cm 4cm},clip,width=0.45\textwidth]{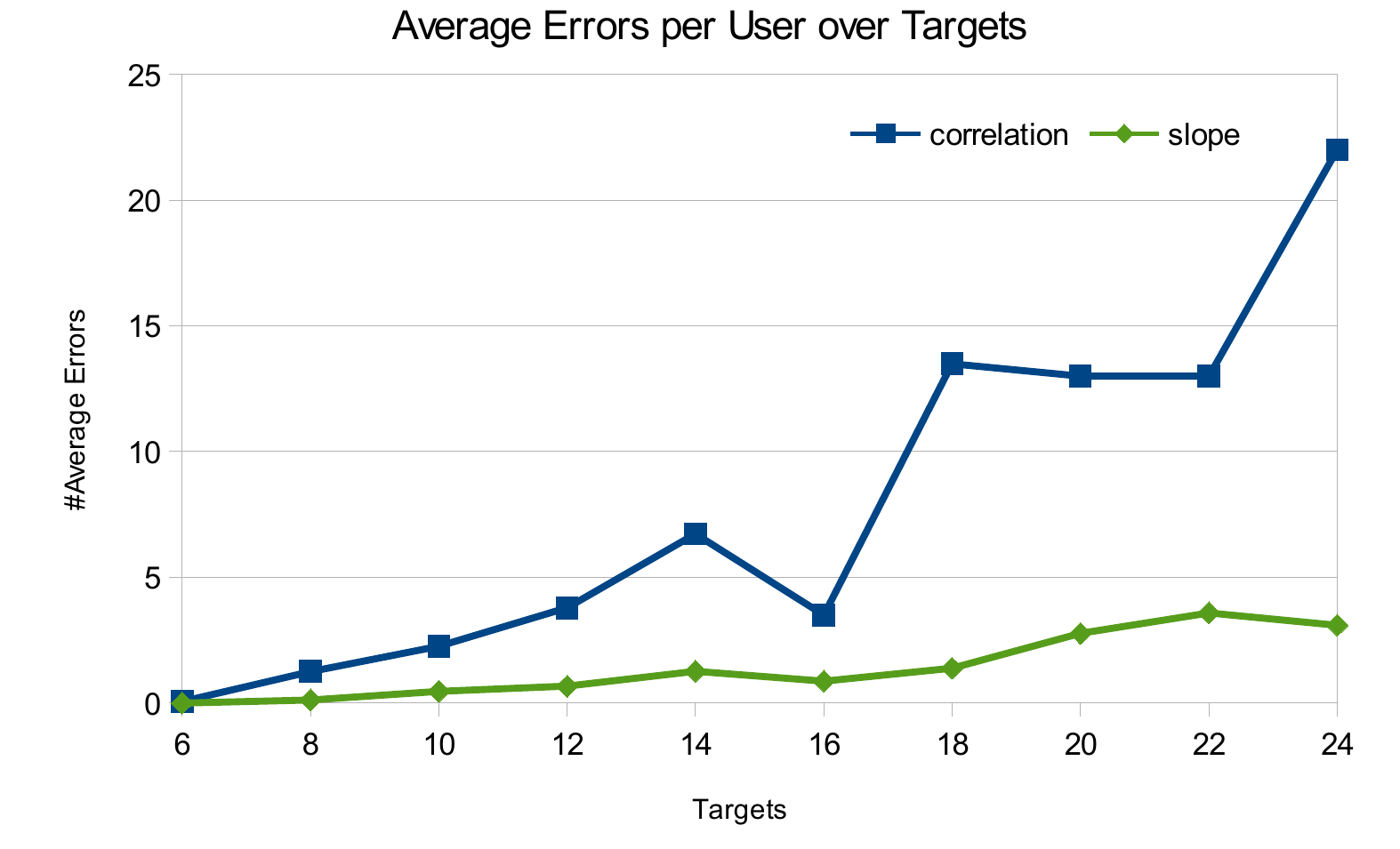}
  \caption{Errors over number of targets. User made consistently fewer errors with the slope method.}
 \label{figure:graph_errors}
\end{figure}

Whenever the participant entered a wrong digit she or he had to cancel the entry with the `cancel'-target. Every entry of the `cancel'-target was counted as error. As seen in Figure \ref{figure:graph_errors}, the average number of errors increases in the presence of more targets, however the increase in errors is sharper in case of the correlation method. For example, while both methods yielded almost no errors at 6 targets across all participants, the mean number of errors at 8 targets was 1.25 and 0.13 for the correlation and slope methods respectively. Similarly, at 24 targets, participants performed 22 errors on average in case of correlation, but only 3 errors on average in case of slope method. Note, Figure \ref{figure:graph_errors} displays an average over the participants who were successful in the respective conditions. 

\subsubsection{Task Completion Time}
\begin{figure} [t]
\centering
  \includegraphics[trim={3.5cm 18cm 1cm 3cm},clip,width=0.5\textwidth]{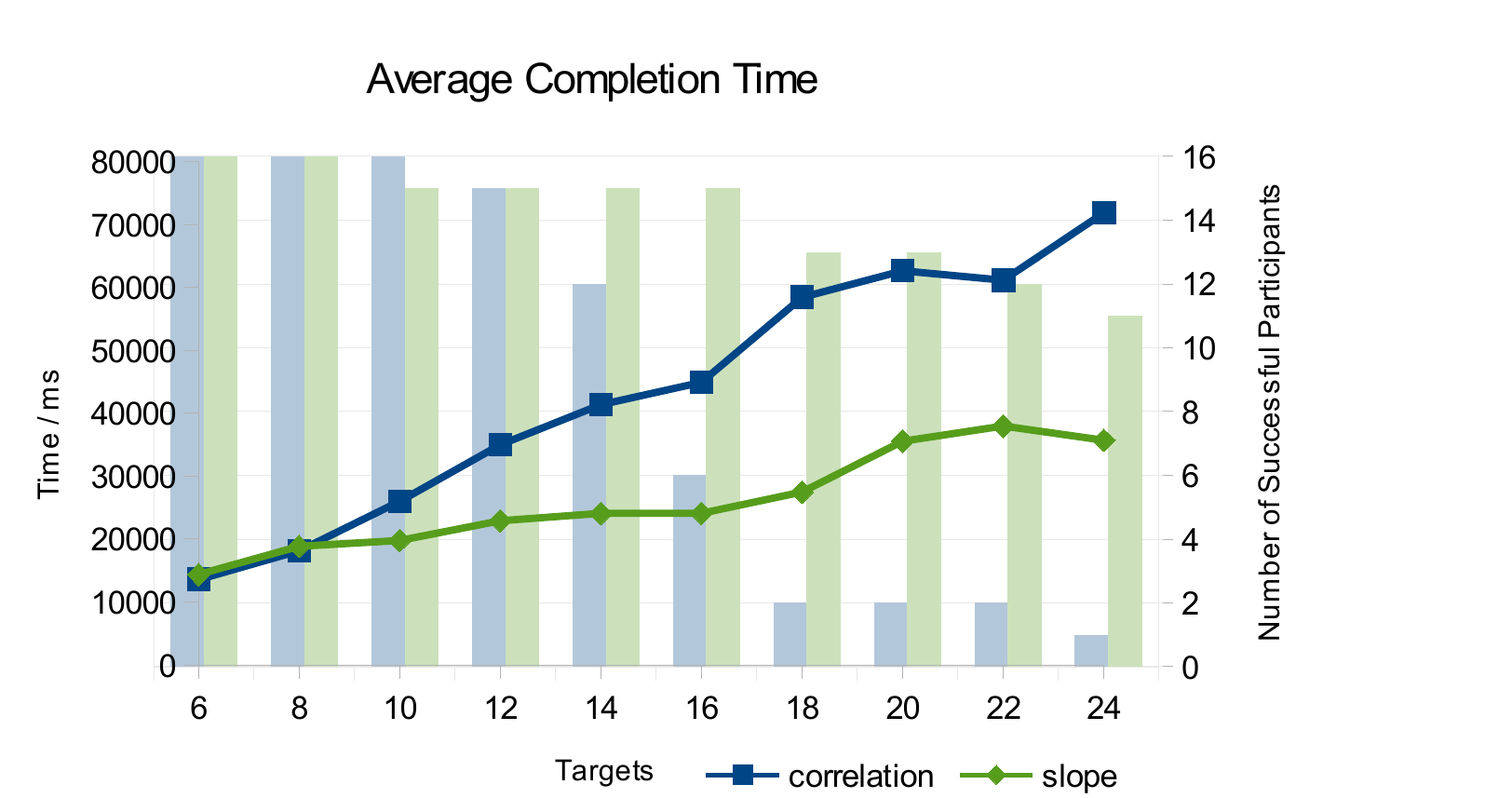}
  \caption{Completion time over number of targets. The slope method was consistently faster than the correlation method. The bars in the background indicate the number of participants successfully completed the task.}
  \vspace{-4mm}
 \label{figure:graph_times}
\end{figure}

We measured the completion time for successfully entering 4 symbols, starting from the moment displaying the symbols, until the moment the fourth symbol was entered. This also includes cancellations. As illustrated by Figure \ref{figure:graph_times}, the average completion time is almost similar across both methods for up to 8 targets, but then increases sharply when using the correlation method compared to when using the slope method. Similar to the errors, successful completion times exclude cases where participants failed to enter the 4 symbols, and hence the average is calculated over a varying number of participants. Completion times are longer for the correlation method. This is mainly due to the many cancellations that participants had to perform. 

\begin{figure} [t]
\centering
\includegraphics[trim={2cm 16cm 3cm 4cm},clip,width=0.45\textwidth]{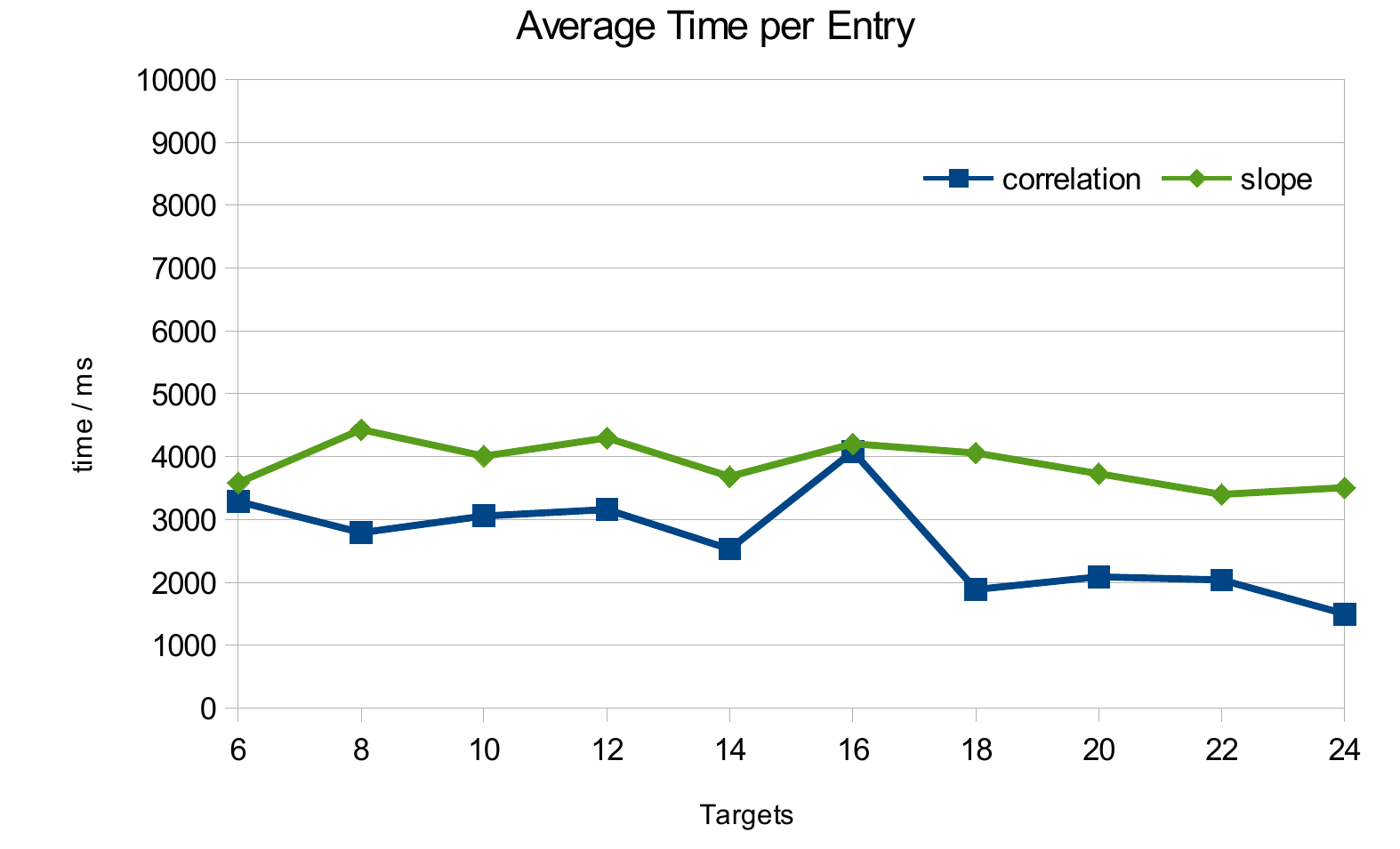}
  \caption{Time per Entry. Participants performed slightly faster on a single entry with the correlation method.}
  \vspace{-3mm}
 \label{figure:graph_timeperentry}
\end{figure}

\subsubsection{Symbol Entry Time}

Unlike the overall completion time which accounts for entering 4 symbols, including the cancellations, this metric reflects the average time it took to select a single entry from the moment the target was gazed at until the moment the target was deemed selected. Figure \ref{figure:graph_timeperentry} shows the times for entering a digit or a cancel operation. 
The slight decrease in selection times could be the results of a learning effect or from the fact that the entry times for the higher target numbers are calculated based on successful participants only, who might as well be well-performing. Entry times did not vary a lot across the detection methods. A Wilcoxon signed rank test showed no evidence of significant effects of detection method on entry time. 

One interesting observation is that the time per entry does not increase with the number of targets. The other interesting observation is that the times for the slope detection method are higher than the times for the correlation detection method (see discussion). This is remarkable as the slope detection uses a smaller minimum signal duration. 

\section{Discussion}
\subsection{Comparing both Methods}

The detection methods studied here depend on different parameters -- the threshold, the data window size, the minimum signal duration, and the smoothing window size. 
A systematic approach with five different values for each parameter would have led to 625 combinations for each detection method.  Hence, we decided to compare the methods using  optimal parameters for each of them. In particular, we used the same correlation value of 0.8 and a data window size of 30 samples as Vidal et al.~\cite{Vidal:2013}. We showed that using a different approach it is possible to almost triple the number of targets. Note, that in our implementation, the correlation method performs even better than in previous work \cite{Vidal:2013}, hence supporting our endeavour to provide a fair comparison. 

\subsection{Understanding the Results}\label{Understanding the Results}

The evaluation yielded significant differences in both methods. We explain and discuss the reasons for these findings. 

\begin{figure} [t]
\centering
  \includegraphics[width=0.30\textwidth]{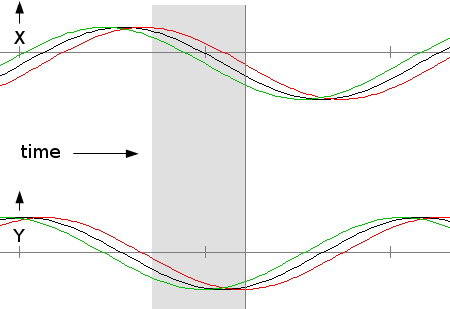}
  \caption{The previous (red) and the next (green) targets are phase shifted by $\pm$2$\pi$/20 against the gaze (black).}
 \label{figure:phase20_part}
\end{figure}

\begin{figure} [t]
\centering
  \includegraphics[width=0.22\textwidth]{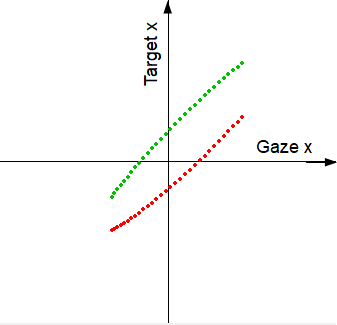}
  \includegraphics[width=0.22\textwidth]{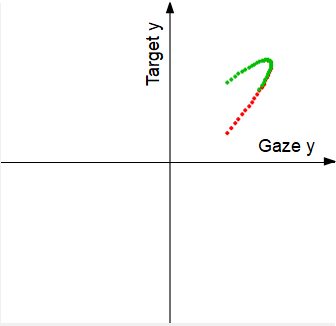}
  \caption{Regression analysis plot for the x-coordinate (left) and the y-coordinate (right).}
 \label{figure:linreg}
\end{figure}

If we assume a perfectly calibrated and accurate eye tracker, and a user whose gaze follows exactly a target on a circle, the x and the y coordinate of the gaze over time would have the shape of a sine and would be shifted $\pi$/2 against each other. 
If there are n targets on the circle, the coordinates of the previous and next target are phase shifted $\pm$2$\pi$/n against the gaze coordinates. The situation for n=20 is depicted in Figure \ref{figure:phase20_part}. The gray area in the figure indicates the current data window. Figure \ref{figure:linreg} shows the regression analysis for the data window in Figure \ref{figure:phase20_part}.

The points all lie on a Lissajous curve which has the shape of an ellipsis. The phase shift affects the eccentricity; the smaller the phase shift the closer the shape is to a diagonal line. The data window size determines the fraction of the ellipsis on which the points lie. 

This allows the influence of the data window size on the detection to be understood. If the data window covers a full cycle, meaning the time for the data window is the time for a target to complete a full circle, the data points form a complete ellipsis. 
In this case, the slope of the regression line and the correlation will be constant over time. If the phase shift is small, both values are close to 1.0. In the depicted case with 20 targets, these values are around 0.95.

With a smaller data window (Figure \ref{figure:phase20_part}), the data points fill only a part of the ellipsis (Figure \ref{figure:linreg}), which moves over time. At the time shown here for the x values, the data points are on an almost straight line and the correlation and the slope are close to 1.0. At the same time, the y values fill the ellipsis tip and the slope of the regression line and the correlation are different from 1.0. As the threshold condition has to be true for the x and y coordinate, this means that there is no positive detection at that moment. However, there are moments in between, where the detection algorithm reports a positive detection.

To get smooth pursuit eye movements, the target speed has to be in a certain range, typically 5--20$^{\circ}$/s.  If we reduce the circle radius and keep the target speed, the cycles per second increase. If we also keep the data window size this means that the data window covers more of the ellipse shown in Figure \ref {figure:linreg}.

\begin{figure} [t]
\centering
  \includegraphics[width=0.40\textwidth]{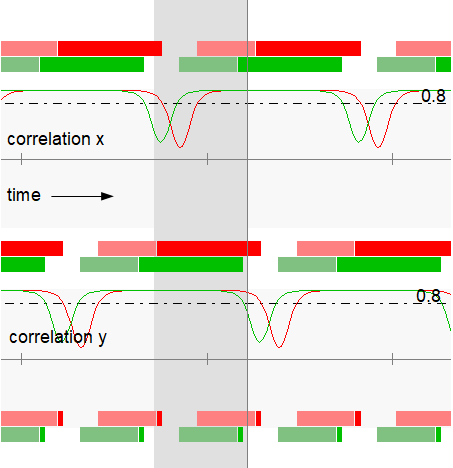}
  \caption{Correlation values for the example in Figure \ref{figure:phase20_part}. The bars indicate a true threshold condition for x (up), y (middle) and both (down). The light color in the bars indicate the minimum signal duration.}
  \vspace{-5mm}
 \label{figure:phase20_cor_part}
\end{figure}

\begin{figure} [t]
\centering
  \includegraphics[width=0.40\textwidth]{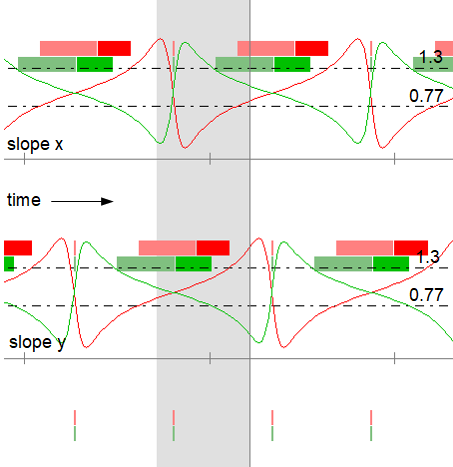}
  \caption{Slope values for the example given in Figure \ref{figure:phase20_part}. The bars have the same meaning as explained in Figure \ref{figure:phase20_cor_part}.}
    \vspace{-4mm}
 \label{figure:phase20_slope_part}
\end{figure}

After understanding the relations of target speed, data window size, and phase difference the question why the slope method performed better remains open. Figure \ref{figure:phase20_cor_part} shows the correlation values for the given example and Figure \ref{figure:phase20_slope_part} shows the values for the slope from the linear regression. The dash-dotted line indicates the thresholds and the bars indicate whether the threshold condition is true. The bars have light color before the minimum signal duration is reached. The lowest bars indicate whether both threshold conditions are true.

The correlation is close to 1.0 most of the time and satisfies the threshold condition (Figure \ref{figure:phase20_cor_part}). The correlation value drops, when the data window covers the ellipsis' tip. As the threshold condition has to be true for the x and the y coordinate, the correlation method signals detection between both drops.

The slope values pass the threshold interval quite quickly and satisfy the threshold condition for a shorter time (see Figure \ref{figure:phase20_slope_part}). The overlap of both signals for the x and y coordinate is minimal. Together with the concept of minimum signal duration, the slope method does not report false positives for 20 targets on a circle (under optimal conditions) while the correlation method does. This is the  reason why the slope method can distinguish more targets on a circle.
On the other hand, this means also that the correlation method detects more easily and more quickly (but at the expense of more false positives). This could explain, why the entry time for the correlation method is slightly shorter (Figure \ref{figure:graph_timeperentry}).

\section{Conclusions and Future Work}

The introduction of a minimum signal duration improves both detection methods, correlation and slope method, as it filters out false positives. 
The new detection algorithm based on the slope of the regression line performs better in separating targets on a circle. 
This does not mean that the slope detection is better in general. 
It seems that the slope detection does not detect true positives as well as the correlation method but creates fewer false positives. 
In the situation of selecting a target from a circle, it is not necessary to have a continuous signal for true positives. 
The first occurrence of a positive signal triggers the entry and possible gaps in the detection signal later do not matter. The property of fewer false positives seems to be more important in our scenario.

The improvement of smooth pursuit detection with the slope method can be either used for increasing the number of targets which offers the user more options or to provide a more robust interface with fewer false positives. We discussed the capabilities of two detection algorithms under idealistic conditions. Future work could try to explain the influence of noise in the gaze data on a theoretical level. 

Directions for future work also  include testing the new detection method in specific application scenarios and with other eye trackers. Researchers could  investigate, how quickly users adapt to such interfaces and whether the need to strongly focus on the target decreases over time. Furthermore, researchers and practitioners could apply and evaluate the slope-based method in domains other than gaze, such as motion matching for body movements \cite{Clarke:2017:RCB:3139486.3130910,Clarke:2016:TCV:2971648.2971714,Clarke:2017:MSS:3126594.3126626}, and mid-air gestures \cite{Carter:2016:PMG:2858036.2858284}.

\balance{}

\bibliographystyle{SIGCHI-Reference-Format}
\bibliography{DialPlate}

\end{document}